\documentclass[preprint,review,12pt,authoryear]{elsarticle}

\usepackage[top=2cm, bottom=2cm, left=2.5cm, right=2.5cm]{geometry}
\usepackage{ragged2e}
\justifying
\usepackage{url}
\usepackage{amssymb}
\usepackage{amsmath}
\usepackage{textcomp} % Load this first to declare standard symbols
\usepackage{gensymb}
\usepackage[inline]{trackchanges}
\usepackage{booktabs}
\usepackage{threeparttable}
\usepackage{multirow}
\usepackage{lmodern}
\usepackage{graphicx}

\journal{Water Research}

\begin{document}

\begin{frontmatter}

\title{Distribution and Transport of Fragmenting Microplastics in a 3D Global Eulerian Model}

\author[label1]{Zih-En Tseng\corref{cor1}} %% Author name
\author[label1]{Yue Wu}
\author[label2]{Chris Ruf}
\author[label3]{Dimitris Menemenlis}
\author[label1]{Yulin Pan}

\cortext[cor1]{Corresponding author}

%% Author affiliation
\affiliation[label1]{organization={Department of Naval Architecture \& Marine engineering, the University of Michigan--Ann Arbor},%Department and Organization
            addressline={1109 Geddes Ave}, 
            city={Ann Arbor},
            postcode={48109}, 
            state={Michigan},
            country={United States}}
            
\affiliation[label2]{organization={Department of Climate and Space Sciences and Engineering, the University of Michigan--Ann Arbor},%Department and Organization
            addressline={2455 Hayward St}, 
            city={Ann Arbor},
            postcode={48109}, 
            state={Michigan},
            country={United States}}

\affiliation[label3]{organization={Moss Landing Marine Laboratories, San Jos\'e State University},%Department and Organization
            addressline={8272 Moss Landing Rd}, 
            city={Moss Landing},
            postcode={95039}, 
            state={California},
            country={United States}}

\begin{abstract}

Fragmentation, the breakage of matter into smaller pieces, is an important mechanism responsible for generating microplastics (MPs). We present the first global three-dimensional Eulerian model that resolves fragmentation alongside MP transport. The evolution of particle size is modeled as a transfer from larger- to smaller-size bins, governed by a fragmentation kinetics framework. Relative to a reference simulation without fragmentation, two distinct effects are identified: (1) the surface concentration field of MPs becomes horizontally dispersed, and (2) MPs sink to depths of $500$~m where the reference simulation shows negligible concentration. The vertical shift can be explained by the loss of buoyancy when particle size decreases, which facilitates horizontal sub-mixed layer transport once the particles sink below $\sim 100$~m depth. Neutrally buoyant particles (with diameter $d<1~\mu\text{m}$) are continuously produced in the ocean by the fragmentation of larger particles and accumulate in the major oceanic gyres. Ultimately, the concentration of these neutrally buoyant MPs peaks at the gyre centers, a behavior that is not captured by prior models. Furthermore, the globally integrated size spectrum exhibits a steepening power-law slope over time that continues to evolve throughout our 25-year simulation. Comparisons with the AOMI Level-3wm observational dataset demonstrate a meaningful improvement in predictive skill relative to previous models: including fragmentation elevates the spatial correlation between modeled and observed surface concentrations from 45\% to 58\%.

\end{abstract}

\begin{highlights}
\item We develop the first global 3D ocean model resolving microplastic fragmentation.
\item Neutrally buoyant fragments accumulate at gyre centers, where they are produced.
\item Fragmentation moves microplastics to 500 m depth and smooths horizontal distribution.
\item Fragmentation reduces concentration variance by up to 7 times in major basins.
\item Fragmentation improves the spatial correlation with observations from 45\% to 58\%.
\end{highlights}


\end{frontmatter}

%%%%%%%%%%%%%%%%%%%%%%%%%%%%%%%%%%%%%%%%%%%%%%%%%%%%%%%%%%%%%%%%%%%%%
%%%%%%%%%%%%%%%%%%%%%%%%%%%%%%%%%%%%%%%%%%%%%%%%%%%%%%%%%%%%%%%%%%%%%
\section{Introduction}
\label{Intro}

%%----------Problem framing-------------------------------------------
Plastic waste is found throughout the global ocean, from the surface mixed layer to the deep sea, and from tropical to polar waters \citep{enders_abundance_2015,ten_hietbrink_nanoplastic_2025}. Recent estimates suggest that up to 27.6 million tons of mismanaged plastic waste are released into the ocean annually \citep{watt_ocean_2021,lau_evaluating_2020,jambeck_plastic_2015}. Once in the ocean, the fragmentation process continuously breaks large debris into microplastics (MPs) through ultraviolet (UV) radiation \citep{pfohl_environmental_2022}, mechanical abrasion \citep{sorasan_ageing_2022}, and degradation \citep{charlesby_polymer_1954}. The resulting microplastic (MP) fragments exhibit a spectrum of sizes, ranging from several nanometers to $5$ millimeters \citep{program_microplastics_2024,ho_comparing_2024}. Studies suggest that these fragments enter the atmosphere through wave-breaking and wind-driven re-suspension, perturbing Earth's radiative balance by scattering and absorbing solar radiation \citep{revell_direct_2021,liu_atmospheric_2026}. The decreasing size of MPs also promotes bio-interaction such as ingestion by animals \citep{jamieson_bioaccumulation_2017} and grazing by zooplankton \citep{richon_faecal_2026}. The broad environmental and biological implications of MP pollution establish it as a major threat to human societies and ecosystems. We therefore need to understand MP transport and distribution in the ocean for risk assessment and informing effective pollution management.

Fragmentation, the process responsible for generating these small MP fragments, is not unique to plastics and has been widely studied across a range of fields and materials, such as polymer degradation \citep{charlesby_polymer_1954}, breakup of liquid droplets \citep{shinnar_liquid_1961,sergeev_droplet_2023}, breakage kinetics in high-energy dry milling operations \citep{capece_grinding_2015}, combustion and the explosion of reactive porous carbon particles \citep{kerstein_fragmentation_1985}, and modal interaction in turbulent fluid flows \citep{chen_turbulent_2020}. Based on the different physical mechanisms responsible for the breakage, the fragmentation processes broadly fall into two classes: linear and non-linear fragmentation, corresponding to the governing equations that describe them. For a linear fragmentation process, in which breakage is driven by an external factor \citep[e.g., UV radiation or mechanical stress;][]{charlesby_polymer_1954,cheng_kinetics_1990,ben_naim_fragmentation_2000}, the breaking rate does not depend on the concentration of particles. While for nonlinear fragmentation, in which the breakage is induced by collision between particles \citep{cheng_kinetics_1990,chen_turbulent_2020}, the breaking rate depends on the concentration of particles and involves high-order terms representing products between concentration fields. For the fragmentation of MPs, polymer degradation driven by external environmental forcing is the dominant mechanism \citep{timar_new_2010,pfohl_environmental_2022,sorasan_ageing_2022}, placing it within the linear fragmentation framework, which this study hinges on.

%%----------Measurements----------------------------------------------
Field measurements directly observe MP concentration in the ocean, primarily through trawler nets and filtering systems, capturing particles $\gtrsim 300\,\mu\text{m}$ and $\gtrsim 10\,\mu\text{m}$, respectively \citep{eriksen_plastic_2014,cozar_plastic_2014,enders_abundance_2015,li_profiling_2020,lindeque_are_2020,pabortsava_high_2020,zhao_distribution_2025,ten_hietbrink_nanoplastic_2025}. These observations have revealed surface garbage patches and the vertical distribution of MPs, and have informed global datasets such as the National Centers for Environmental Information \citep[NCEI;][]{nyadjro_noaa_2023} and Atlas of Ocean Microplastics \citep[AOMI;][]{isobe_multilevel_2021}. However, direct measurements remain constrained by the under-representation of particles below the sampling threshold and by limited spatial and temporal coverage. Remote-sensing techniques \citep{jones-williams_remote_2021}, such as the Cyclone Global Navigation Satellite System \citep[CYGNSS][]{Maddy_Chris_2022,cygnss_team_cygnss_2024}, offer more extensive and uniform coverage. Still, the results rely on indirect retrieval algorithms sensitive to conditions such as air-sea heat flux, and must be interpreted with caution.

%%----------Models----------------------------------------------------

Numerical modeling complements these observations, broadly through Lagrangian and Eulerian approaches, with the former tracking individual particle trajectories and the latter resolving a particle concentration field. The Eulerian approach sacrifices individual-particle history in return for several advantages---most notably for the present study, a computational cost that does not scale with the number of particles, which is significantly increased by fragmentation \citep{tseng_distribution_2025}. As both classes of models advance, so does our understanding of the transport of MPs. Early two-dimensional (2D) Lagrangian models revealed that buoyant particles accumulate within subtropical gyres \citep{lebreton_numerical_2012,chenillat_fate_2021}. The subsequent three-dimensional (3D) models resolved the vertical motion of MPs and distinguished particles by their properties such as density and size \citep[e.g.,][]{lobelle_global_2021,mountford_eulerian_2019,richon_zooplankton_2022,richon_faecal_2026,tseng_effect_2026}. In particular, the recent work by \citet{tseng_distribution_2025} has established that particle size is a critical factor governing the transport and distribution of low density MPs (i.e., with density lower than seawater). Specifically, larger particles (with diameter $\gtrsim 10 \,\mu\text{m}$) aggregate within the mixed layer (ML) in subtropical gyres, while particles small enough (with diameter $\lesssim 1 \mu\text{m}$) can be treated as neutrally buoyant, and are instead transported deeper into the interior water rather than forming surface garbage patches in the gyres.

%%----------Motivate-fragmentation------------------------------------
Most models summarized above assume that the particle size remains constant in time during their transport. This assumption prevents such models from capturing fragmentation and brings some problems in the models. For example, all small particles would be generated only along the coast. However, smaller plastic fragments can be generated from the degradation of larger debris, which occurs ubiquitously instead of only at the coast. This bias prevents existing models from capturing how particle size evolves during transport and where smaller MPs are generated. Since particle size governs their transport pathway, this in turn limits the ability of existing models to predict the global distribution of MPs.

%%----------This-study------------------------------------------------
Motivated by all of the above reasons, we develop the first 3D global Eulerian model that directly resolves fragmentation along with the transport of MPs. After a 25-year simulation, the distribution of fragmenting MPs exhibits several differences compared to the reference distribution of fixed-size MPs: (1) In terms of surface concentration, fragmentation disperses the surface garbage patches and weakens the horizontal gradient. (2) In terms of vertical distribution, MP fragments are identified below the mixed layer, reaching depths of $500$ meters. In particular, we find that the concentration of neutrally buoyant particles peaks at gyre centers, which cannot be predicted by earlier models. This is because neutrally buoyant MPs are generated in situ by the fragmentation of larger particles, which quickly accumulate (in $<10$ years, compared to the full 25-year duration of the simulation) within subtropical gyres. (3) Consistent with the fragmentation theories, the simulated size spectrum of MPs reproduces the power-law form with a slope that is still evolving toward the steady state. (4) Finally, incorporating fragmentation improves the model-to-observation agreement, raising the spatial correlation with observed surface concentrations \citep{isobe_multilevel_2021} from 45\% to 58\%.

This paper is structured as follows. Section~\ref{sec:methodology} presents the modeling details; Section~\ref{sec:results} discusses the resulting distributions and size-spectrum evolution; and Section~\ref{sec:conclusion} concludes the work with final remarks.

%%%%%%%%%%%%%%%%%%%%%%%%%%%%%%%%%%%%%%%%%%%%%%%%%%%%%%%%%%%%%%%%%%%%%
%%%%%%%%%%%%%%%%%%%%%%%%%%%%%%%%%%%%%%%%%%%%%%%%%%%%%%%%%%%%%%%%%%%%%
\section{Methodology}
\label{sec:methodology}

%%--------------------------------------------------------------------
%%--------------------------------------------------------------------
\subsection{Fragmentation equation}
\label{sec:methods_fragmentation}

\begin{figure}[!ht]
\centering
\includegraphics[width=.7\textwidth]{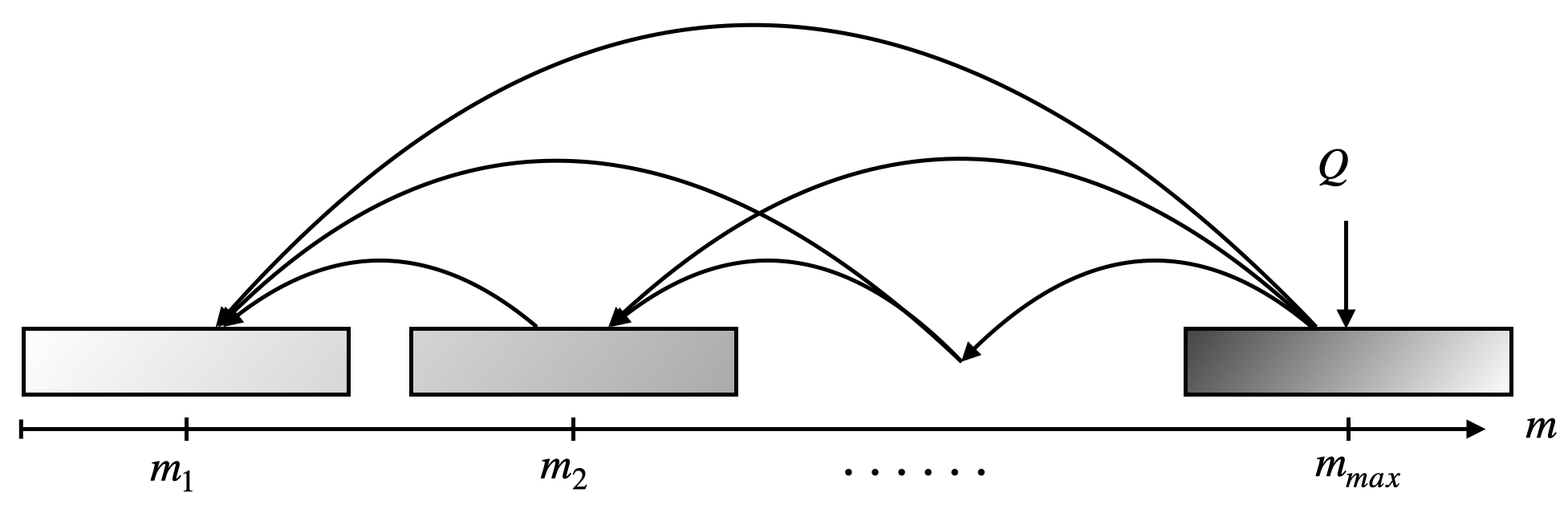}
\caption{Illustration of the fragmentation loss/gain mechanism. The production of particles involves the contributions from the full range of larger masses.}
\label{fig:frag_schematic}
\end{figure}

We describe the broadband distribution of MPs using the mass density spectrum of their number concentration, $\tau(m,t)$ (with unit $\#\,\text{m}^{-3}\,\text{g}^{-1}$). The evolution of $\tau(m,t)$ under linear fragmentation with a steady source is described in \citet{ben_naim_fragmentation_2000}
%%%%%%%%%%%%%%
\begin{equation}
\frac{d\tau(m,t)}{dt} = -a(m)\,\tau(m,t) + \int_{m}^{\infty} a(\eta)\,b(m|\eta)\, \tau(\eta,t)\,d\eta + Q\,\delta(m-m_\text{max}),
\label{eq:fragmentation}
\end{equation}
%%%%%%%%%%%%%%
where $m_\text{max} \approx 5.9 \times 10^{-8}$ g corresponds to the mass of a spherical particle with diameter $d=50\,\mu\text{m}$ and particle density $\rho_p=900$ kg m$^{-3}$. The first term on the right-hand side represents the loss due to the breaking of particles with mass $m$ (g) into smaller particles. The mass-dependent fragmentation rate, $a(m) = a_0 m^\lambda$ with $\lambda > 0$, characterizes the non-shattering (larger particles fragmenting faster) behavior of MPs \citep{cheng_kinetics_1990,banasiak_shattering_2006}. We choose $a_0 \approx 9.66 \times 10^{-8}$ s$^{-1}$ g$^{-\lambda}$ and $\lambda = 0.2$, corresponding to a fragmentation rate of $0.9 \%$ per month \citep{gerritse_fragmentation_2020} for particles with $m=m_\text{max}$. The second term is a gain term, where the breaking of larger particles with mass $\eta$ produces the smaller particles with mass $m$ conditioned on the density function
%%%%%%%%%%%%%%
\begin{equation}
    b(m|\eta) = \frac{\nu+2}{\eta}\!\left(\frac{m}{\eta}\right)^{\!\nu},
\end{equation}
%%%%%%%%%%%%%%
with $m < \eta$ and $\nu \in (-2,0]$ \citep{cheng_kinetics_1990}. Particles with mass $m$ can be produced by all larger particles with mass $\eta > m$, so that this gain rate involves integrating the productions from mass $m$ to infinity. Here $\nu = -5/3$ is chosen based on the observation by \citet{cozar_plastic_2014} and the experiment by \citet{sorasan_ageing_2022}. The third term is a source term, with $\delta$ being the Dirac delta function, representing the release of particles with mass $m=m_\text{max}$ at rate $Q$. The linear fragmentation process is illustrated in Figure~\ref{fig:frag_schematic}.

%%%%%%%%%%%%%%
%%%%%%%%%%%%%%

The steady-state solution to Eq.~\eqref{eq:fragmentation} has been studied by \citet{cheng_kinetics_1990}, which analyzed the source-free case, and \citet{ben_naim_fragmentation_2000}, which introduced a steady source but restricted their analysis to a specific combination of parameters ($\lambda=1$, $\nu=0$). Still, neither of them provides a closed-form steady-state solution for general $\lambda$ and $\nu$. Here we derive such a solution, which takes the power-law form
%%%%%%%%%%%%%%
\begin{equation}
\lim_{t \to \infty} \tau(m,t) = c_L \left(\frac{m}{m_\text{max}}\right)^{-(\lambda+2)}, \quad m \leq m_\text{max},
\label{eq:steady-state-solution}
\end{equation}
%%%%%%%%%%%%%%
with $c_L \equiv \lim_{t \to \infty} \tau(m_\text{max},t) = Q/(a_0 m_\text{max}^\lambda)$. The exponent $-(\lambda+2)$ depends only on the fragmentation rate exponent $\lambda$ and is independent of the density function parameter $\nu$ related to $b(m,\eta)$. A detailed derivation and a series of numerical tests to confirm the solution are provided in \ref{append:steady-state}.

%%--------------------------------------------------------------------
%%--------------------------------------------------------------------
\subsection{Transport equation with fragmentation}
\label{sec:methods_transport}

We incorporate the fragmentation equation, Eq.~\eqref{eq:fragmentation}, into the MIT general circulation model \citep[MITgcm;][]{marshall_hydrostatic_1997,marshall_finitevolume_1997} to study its effect on the global distribution of MPs. Adding the fragmentation terms to the governing transport equation \citep{tseng_distribution_2025} gives
%%%%%%%%%%%%%%
\begin{equation}
\begin{aligned}
    \frac{\partial \tau(\vec{x},m,t)}{\partial t} + \nabla \cdot (\tau \mathbf{u}) +\partial_z (\tau w_r) =\nabla \cdot (K \nabla \tau) + Q(\vec{x}) \,\delta(m-m_\text{max}) \\
    -a(m)\,\tau(\vec{x},m,t) + \int_{m}^{\infty} a(\eta)\,b(m|\eta)\, \tau(\vec{x},\eta,t)\,d\eta,
    \label{eq:adv-diff}
\end{aligned}
\end{equation}
%%%%%%%%%%%%%%
where $\mathbf{u}$ and $K$ are, respectively, the ocean current velocity vector and diffusivity tensor from the Estimating the Circulation and Climate of the Ocean Version 4 Release 5 (ECCO V4r5) global ocean state estimate \citep{forget_ecco_2015,ecco_consortium_synopsis_2024}, and $w_r$ is the particle terminal velocity that depends on the size and shape of the grains. Since the particles we consider are small enough, we may safely assume that they are spherical in the low-Reynolds-number limit \citep{dey_terminal_2019}. Thus, the particle mass $m$ is related to diameter $d$ through $m = \dfrac{\pi}{6}\rho_pd^3$, and the terminal velocity $w_r$ is
%%%%%%%%%%%%%%
\begin{equation}
    w_r = \dfrac{g d^2 (\rho_w-\rho_p)}{18 \mu},
    \label{eq:terminal_velocity}
\end{equation}
%%%%%%%%%%%%%%
where $g$ is gravitational acceleration, $\rho_w$ is water density, and $\mu$ is the dynamic viscosity. We only release large particles with $m=m_\text{max}$, and $Q(\vec{x})$ is the coastal release rate according to \citet{jambeck_plastic_2015}, \citet{mountford_eulerian_2019}, and \citet{tseng_distribution_2025}. 

In the numerical model, particle masses are discretized into $N=10$ logarithmically spaced bins, spanning from $4.7 \times 10^{-13}$ g to $1.0 \times 10^{-7}$ g (corresponding to diameters from $1\,\mu\text{m}$ to $60\,\mu\text{m}$; Table~\ref{tab:bins}). This logarithmic spacing is chosen for two reasons. First, \citet{tseng_distribution_2025} identified a transition of particles transport behavior for particles between 1~$\mu\text{m}$ and 10~$\mu\text{m}$, shifting from positively buoyant ($d\geq$10~$\mu\text{m}$) to neutrally buoyant ($d\leq$1~$\mu\text{m}$). A finer bin resolution within this diameter range is needed to properly resolve this transition. Second, narrow bin widths at the lower end mitigate artificial over-accumulation of particles within the smallest-size bin, which would otherwise occur if we used one wide bin to represent the particles converging toward $m\to0$.

The choice of $N=10$ bins reflects a trade-off between the desired resolution and computational cost (Table~\ref{tab:bins}). The simulation spans 25 years, a duration primarily constrained by the availability of the ECCO V4r5 ocean state estimate, and was performed on Bridges-2 \citep{psc_bridges2_2024} at the Pittsburgh Supercomputing Center, using 96 cores for 2 days of wall-time per simulation with a 20 minute time step. During our preliminary simulation (\ref{append:steady-state-numerical}), we have checked that our fragmentation system captures the previously derived steady state without source \citep{cheng_kinetics_1990} and our newly derived steady state with source. We find that $N=10$ bins are sufficient for the purposes of this study. The discretization of Eq.~\eqref{eq:adv-diff}, including the discrete transfer rate between bins, is presented with more detail in \ref{append:discretization}.

%%%%%%%%%%%%%%
\begin{table}[!ht]
\centering
\caption{Definition of the $N=10$ mass bins, showing their diameter range and the corresponding mass range, bridged via the spherical-particle assumption.}
\label{tab:bins}
\begin{tabular}{c cc c}
\toprule
Bin $i$ & Diameter range ($\mu\text{m}$) & Lower mass bound $L_i$ (g) & Upper mass bound $R_i$ (g) \\
\midrule
1  & (0.000, 1.000]   & 0                      & $4.712\times10^{-13}$ \\
2  & (1.000, 1.576]   & $4.712\times10^{-13}$  & $1.845\times10^{-12}$ \\
3  & (1.576, 2.484]   & $1.845\times10^{-12}$  & $7.222\times10^{-12}$ \\
4  & (2.484, 3.915]   & $7.222\times10^{-12}$  & $2.827\times10^{-11}$ \\
5  & (3.915, 6.170]   & $2.827\times10^{-11}$  & $1.107\times10^{-10}$ \\
6  & (6.170, 9.724]   & $1.107\times10^{-10}$  & $4.333\times10^{-10}$ \\
7  & (9.724, 15.326]  & $4.333\times10^{-10}$  & $1.696\times10^{-9}$ \\
8  & (15.326, 24.155] & $1.696\times10^{-9}$   & $6.641\times10^{-9}$ \\
9  & (24.155, 38.070] & $6.641\times10^{-9}$   & $2.600\times10^{-8}$ \\
10 & (38.070, 60.000] & $2.600\times10^{-8}$   & $1.018\times10^{-7}$ \\
\bottomrule
\end{tabular}
\end{table}
%%%%%%%%%%%%%%

% The mass concentration of MPs from each mass bin $i$, and the total mass concentration of all particles combined, are defined respectively as $M_i(\vec{x},t) = \int_{L_i}^{R_i} m\,\tau(\vec{x},m,t)\,dm$ and $M(\vec{x},t) = \sum_i^N M_i(\vec{x},t)$; we will revisit these definitions in Section~\ref{sec:results} immediately before each quantity is shown on a map.

%%%%%%%%%%%%%%%%%%%%%%%%%%%%%%%%%%%%%%%%%%%%%%%%%%%%%%%%%%%%%%%%%%%%%
%%%%%%%%%%%%%%%%%%%%%%%%%%%%%%%%%%%%%%%%%%%%%%%%%%%%%%%%%%%%%%%%%%%%%
\section{Results and discussion}
\label{sec:results}

The results are organized as follows: we begin with Sec.~\ref{sec:results_collective} presenting the collective concentration of MPs across all mass bins combined, to showcase how fragmentation reshapes the global MP distribution. Next, in Sec.~\ref{sec:results_individual} the contribution from individual particle size bins is detailed, showing the distinct spatial patterns of large, intermediate, and small particles. Then, Sec.~\ref{sec:results_spectrum} and Sec.~\ref{sec:results_3d} show the global and local mass density spectra, respectively, with the analysis performed in Sec.~\ref{sec:methods_fragmentation} as a guidance. Finally, Sec.~\ref{sec:results_validation} closes the section with a quantitative evaluation against observational data from \citet{isobe_multilevel_2021}.

%%--------------------------------------------------------------------
%%--------------------------------------------------------------------
\subsection{Collective mass concentration of microplastics across all bins}
\label{sec:results_collective}

\begin{figure}[!ht]
\centering
\includegraphics[width=1.0\textwidth]{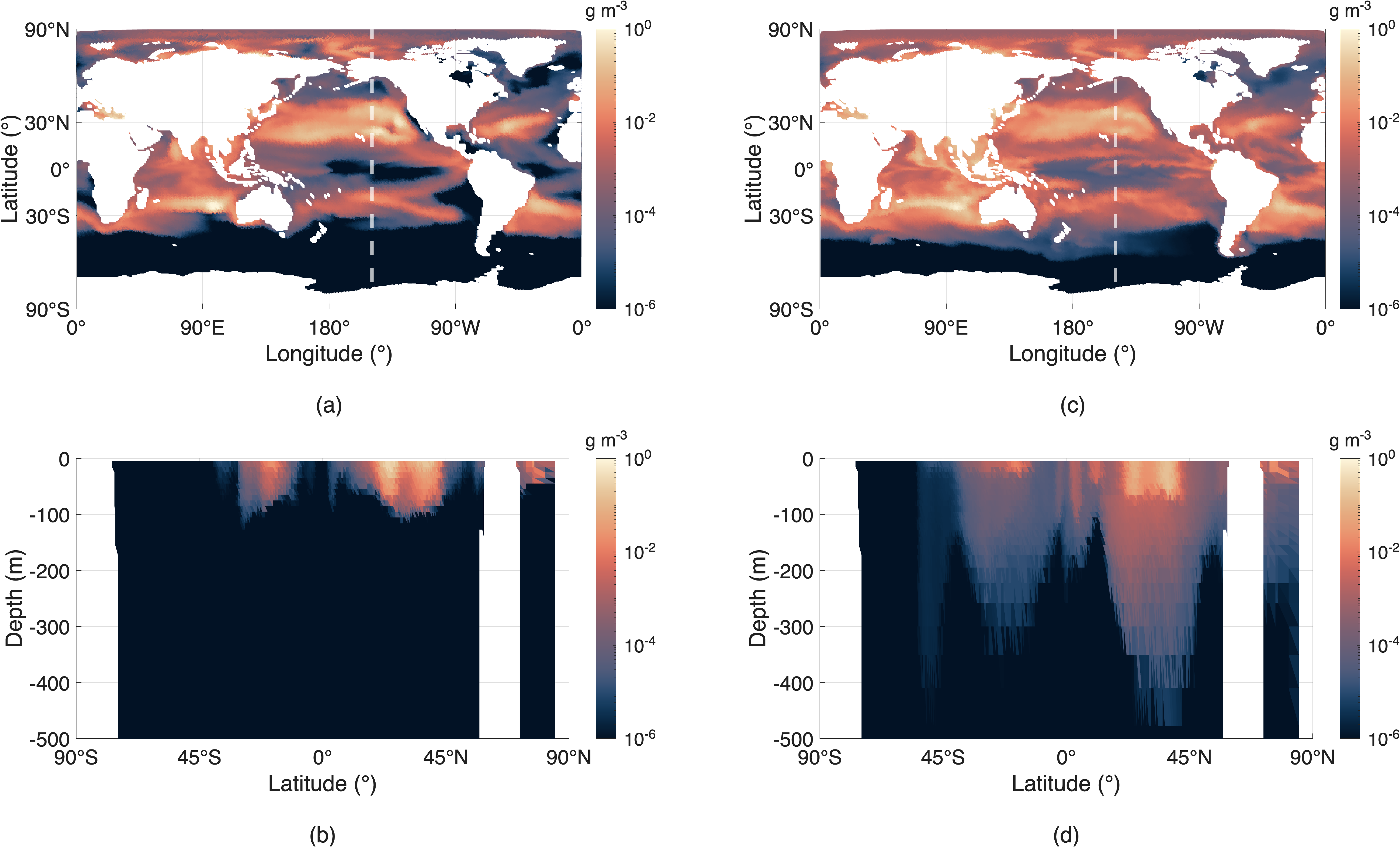}
\caption{(a) Mass concentration $M$ of microplastics at the global sea surface without fragmentation, with a white dashed line denoting $150\degree$W. (b) Vertical slice of $M$ without fragmentation across $150\degree$W. (c) and (d), the same as (a) and (b) but with fragmentation.}
\label{fig:collective_surface}
\end{figure}

We first define the collective mass concentration $M(\vec{x},t)$, summing the contribution of all particles over the entire mass range
%%%%%%%%%%%%%%
\begin{equation}
    M(\vec{x},t) = \int_{0}^{R_N} m \, \tau(\vec{x},m,t) \, dm.
\end{equation}
%%%%%%%%%%%%%%
To provide a baseline for the following comparison, we have performed a reference simulation considering no fragmentation, where all the particles released have $m=m_\text{max}$, do not breakdown (effectively with $a(m)\equiv 0$ so that no smaller particles are generated), and are persistently transported for a long time. A comparison between the modeled total mass concentration $M(\vec{x},t)$ of MPs without fragmentation (fixed-size particles) and with fragmentation (fragmenting particles) is shown in Figure~\ref{fig:collective_surface}.

%%%%%%%%%%%%%%
%%%%%%%%%%%%%%

For fixed-size particles, their distribution pattern is the same with the earlier studies of positively buoyant particles. On the sea surface (Figure~\ref{fig:collective_surface}a), garbage patches can be clearly identified in the Indian, South Pacific, North Pacific, South Atlantic, North Atlantic, and the Arctic oceans \citep{mountford_eulerian_2019,richon_zooplankton_2022,tseng_distribution_2025}. On the vertical slice taken at $150\degree$W (Figure~\ref{fig:collective_surface}b), particles aggregate in the mixed layer within $\lesssim$ 100 m depth \citep{richon_zooplankton_2022,tseng_distribution_2025}.

When fragmentation is considered, two effects emerge in the horizontal and vertical distributions. First, the surface concentration field of fragmenting particles exhibits lower horizontal gradients relative to the fixed-size case, indicating a horizontal smoothing (Figure~\ref{fig:collective_surface}c). Second, signatures of plastic fragments can be identified at 500 m depth, indicating a vertical deepening (Figure~\ref{fig:collective_surface}d). 

The vertical deepening is explained by the loss of buoyancy as the particle size continuously decreases due to fragmentation. Compared to fixed-size particles, the small particle fragments have a lower terminal velocity, which offers less resistance to turbulent vertical mixing. Thus, fragmentation-generated smaller particles are more easily entrained below the mixed layer \citep{mountford_eulerian_2019,tseng_distribution_2025}. The horizontal smoothing can then be explained by the sub-mixed layer transport \citep[][see their Figure 10]{tseng_effect_2026}, which results in a net flux of particles outward from the gyre centers and horizontally disperses the garbage patches. Subsequently, some of the dispersed particles are re-entrained in the surface mixed layer.

%%--------------------------------------------------------------------
%%--------------------------------------------------------------------
\subsection{Individual mass concentration of microplastics in selected bins}
\label{sec:results_individual}

We now examine the contribution to the collective concentration of MPs from individual mass bins
%%%%%%%%%%%%%%
\begin{equation}
    M_i(\vec{x},t) = \int_{L_i}^{R_i} m \, \tau(\vec{x},m,t) \, dm.
\end{equation}
%%%%%%%%%%%%%%
Specifically, we focus on three mass bins: (1) the largest-size bin ($i=10$), corresponding to particles with diameter $d \in (38 \, \mu\text{m}$, $60 \, \mu\text{m}]$; (2) an intermediate-size bin ($i=6$), corresponding to particles with diameter $d \in (6.1 \, \mu\text{m}$, $9.7 \, \mu\text{m}]$; and (3) the smallest-size bin ($i=1$), corresponding to particles with diameter $d \in (0 \, \mu\text{m}$, $1 \, \mu\text{m}]$.

%%%%%%%%%%%%%%
\begin{figure}[!ht]
\centering
\includegraphics[width=1.0\textwidth]{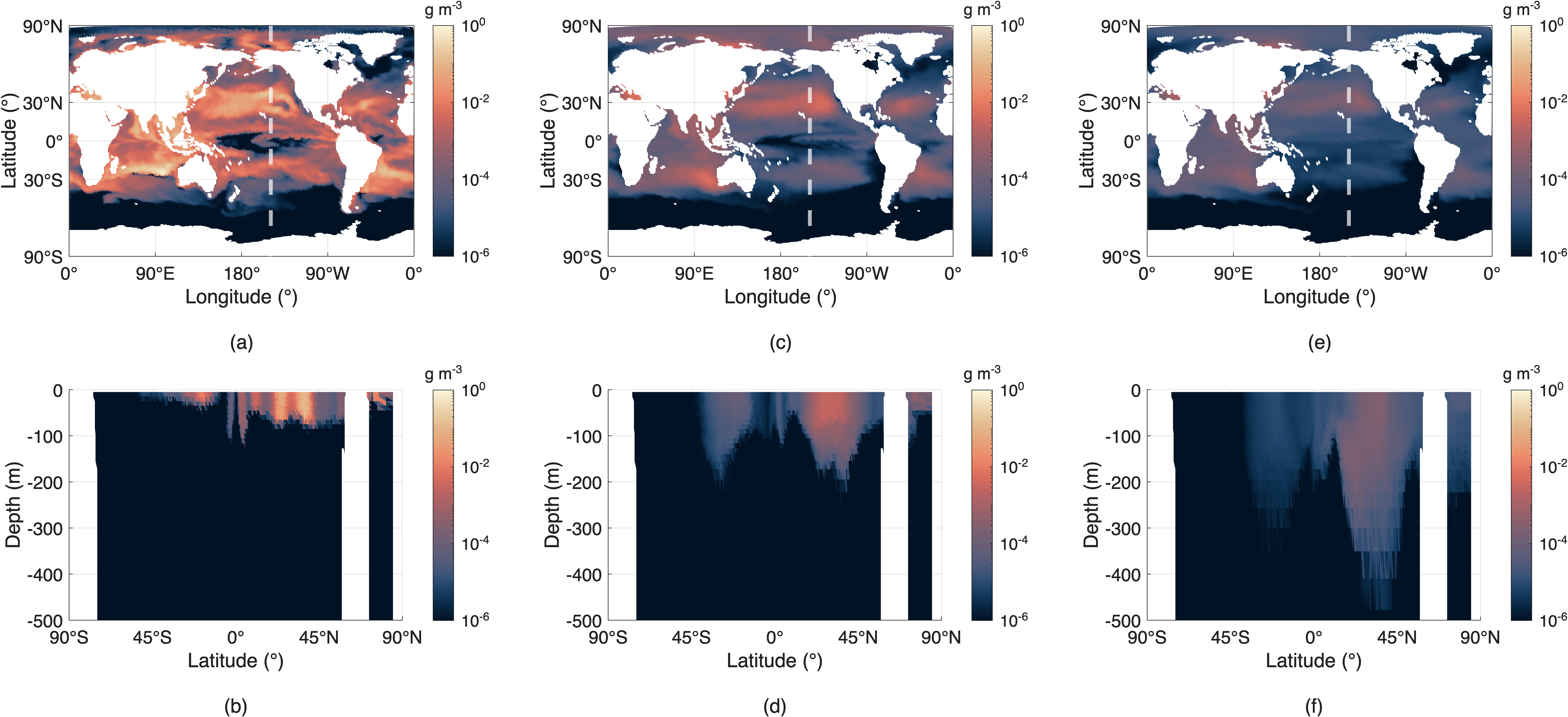}
\caption{(a,\,c,\,e) Sea-surface mass concentration for bins $M_{10}$, $M_{6}$, and $M_{1}$ that correspond to bin center mass $m=6.39\times10^{-8}\,\text{g},\,2.72\times10^{-10}\,\text{g}$, and $2.36\times10^{-13}\,\text{g}$ and diameter $d=51.4\,\mu\text{m},\,8.33\,\mu\text{m}$, and $0.79\,\mu\text{m}$, respectively. White dashed lines denote $150\degree$W. (b,\,d,\,f) Vertical slices of mass concentration across $150\degree$W for $M_{10}$, $M_{6}$, and $M_{1}$.}
\label{fig:individuals}
\end{figure}
%%%%%%%%%%%%%%

For particles from the largest-size bin, their surface concentration $M_{10}$ reproduces the well-documented garbage patches located at the centers of the subtropical gyres (Figure~\ref{fig:individuals}a). Contributing $\sim 63 \%$ to the total mass of particles in the simulation domain, the surface pattern of $M_{10}$ (Figure~\ref{fig:individuals}a) thus dominates that of the total concentration $M$ shown in Figure~\ref{fig:collective_surface}c. In the vertical direction (Figure~\ref{fig:individuals}b), the strong buoyancy of these particles confines them within the ocean surface mixed layer.

Next, the concentration $M_{6}$ of particles from an intermediate-size bin is shown on the sea surface (Figure~\ref{fig:individuals}c) and across a vertical transect (Figure~\ref{fig:individuals}d). Earlier studies have shown that $10\,\mu\text{m}$ MPs possess sufficient buoyancy to aggregate within the mixed layer and form surface patches \citep{tseng_distribution_2025}. The particle fragments in this mass bin have diameters close to the $10\,\mu\text{m}$ that was previously considered. The formation of surface garbage patch we see in Figure~\ref{fig:individuals}c and the accumulation of particles in the mixed layer in Figure~\ref{fig:individuals}d are thus as expected.

The concentration $M_{1}$ of the smallest-size particles is then shown at the sea surface and on a vertical slice in Figures~\ref{fig:individuals}e and \ref{fig:individuals}f. At the sea surface (Figure~\ref{fig:individuals}e), the concentration $M_{1}$ is found to peak at the gyre centers, forming garbage patches. In the vertical direction (Figure~\ref{fig:individuals}f), particles with a size $d \lesssim 1\,\mu\text{m}$ have negligible terminal velocity and can be identified at $\sim500$ m depth. The formation of surface patches in $M_{1}$ was not predicted by the previous study of neutrally buoyant particles \citep[the $d=1\,\mu\text{m}$ case by][]{tseng_distribution_2025,mountford_eulerian_2019}, where the concentration was found to peak near coastlines. The contrast is due to the different treatment of the way these particles are generated: \citet{tseng_distribution_2025} problematically assume that all particles already become neutrally buoyant fragments before they are released along the coast, while the current study more realistically considers neutrally buoyant particles to be generated by fragmentation during the transport of larger-size particles. Both studies indicate that the distribution of neutrally buoyant particles is production-driven, i.e., it is highly correlated to the sites of their generation. In the current study, the surface patch pattern of $M_{10}$ forms in $<10$ years, which subsequently serves as a source of $M_{1}$ throughout the 25-year simulation, determining the $M_{1}$ surface pattern shown in Figure~\ref{fig:individuals}e.

%%--------------------------------------------------------------------
%%--------------------------------------------------------------------
\subsection{Global spectrum of number concentration}
\label{sec:results_spectrum}

\begin{figure}[!ht]
\centering
\includegraphics[width=0.7\textwidth]{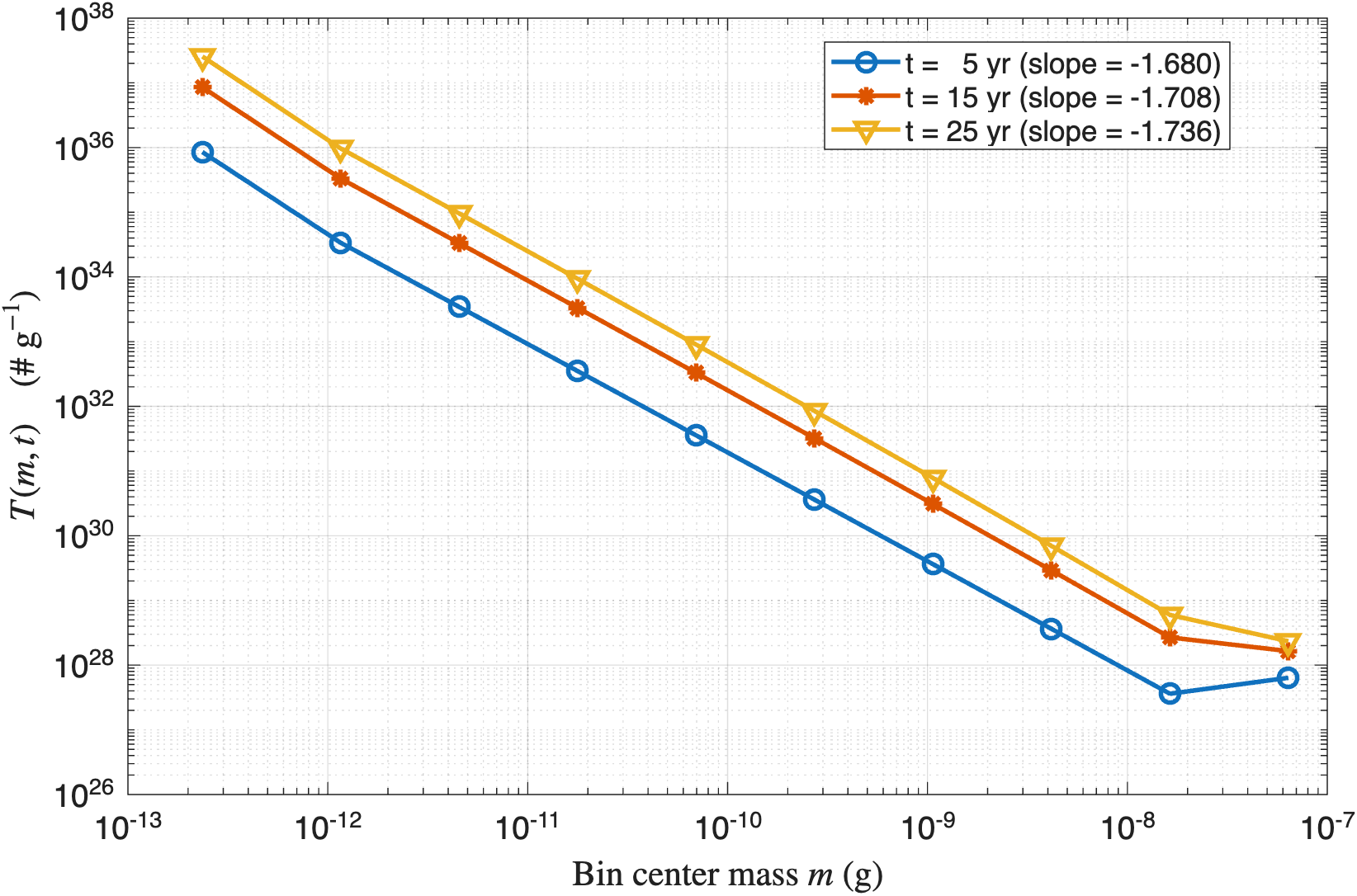}
\caption{The globally integrated spectrum $T(m,t)$ and its best-fit slope, shown for several times during the simulation.}
\label{fig:spectrum_evolution}
\end{figure}

The globally integrated mass density spectrum of MP number concentration 
%%%%%%%%%%%%%%
\begin{equation}
    T(m,t) = \iiint \tau(\vec{x},m,t) \, d\vec{x}, 
\end{equation}
%%%%%%%%%%%%%%
is shown for three instances of time during the simulation in Figure~\ref{fig:spectrum_evolution}. The solution exhibits a power-law form throughout the simulation, consistent with the prediction of linear fragmentation theory \citep{ben_naim_fragmentation_2000}. In terms of the spectrum's temporal evolution, the power-law slope continues to steepen over the course of the simulation and has not yet reached the steady state. This steepening follows from the fragmentation mechanism, which persistently converts mass from large particles into small particles. Thus, the relative concentration of small particles continues to grow at the expense of large particles, and the slope steepens accordingly. The 25-year simulation has not reached the steady-state slope of $-(\lambda+2)=-2.2$. In fact, our simulation of the fragmentation system alone (\ref{append:steady-state-numerical}) shows that the slope takes at least 200 years to reach the steady state exhibiting the slope of $-2.2$.

%%%%%%%%%%%%%%
%%%%%%%%%%%%%%

% For a more intuitive comparison, we compute the first-order moment of the spectrum $m\, T(m,t)$, which converts the number concentration of MPs into the mass concentration. The moment is then transformed from the $m$-space onto the $d$-space through the Jacobian $\partial m/\partial d$, meaning that we converts the mass density spectrum into a diameter density spectrum. The result is shown in Figure~\ref{fig:spectrum_evolution}b, where the slope of the diameter-density spectrum is flat, with a slope close to zero, at the small-size end. This means that if the MPs are divided into uniform diameter-bins, we should expect the small-size bins to cover a similar total mass of particles across the bins.

%%--------------------------------------------------------------------
%%--------------------------------------------------------------------
\subsection{Local spectrum of number concentration}
\label{sec:results_3d}

\begin{figure}[!ht]
\centering
\includegraphics[width=1.0\textwidth]{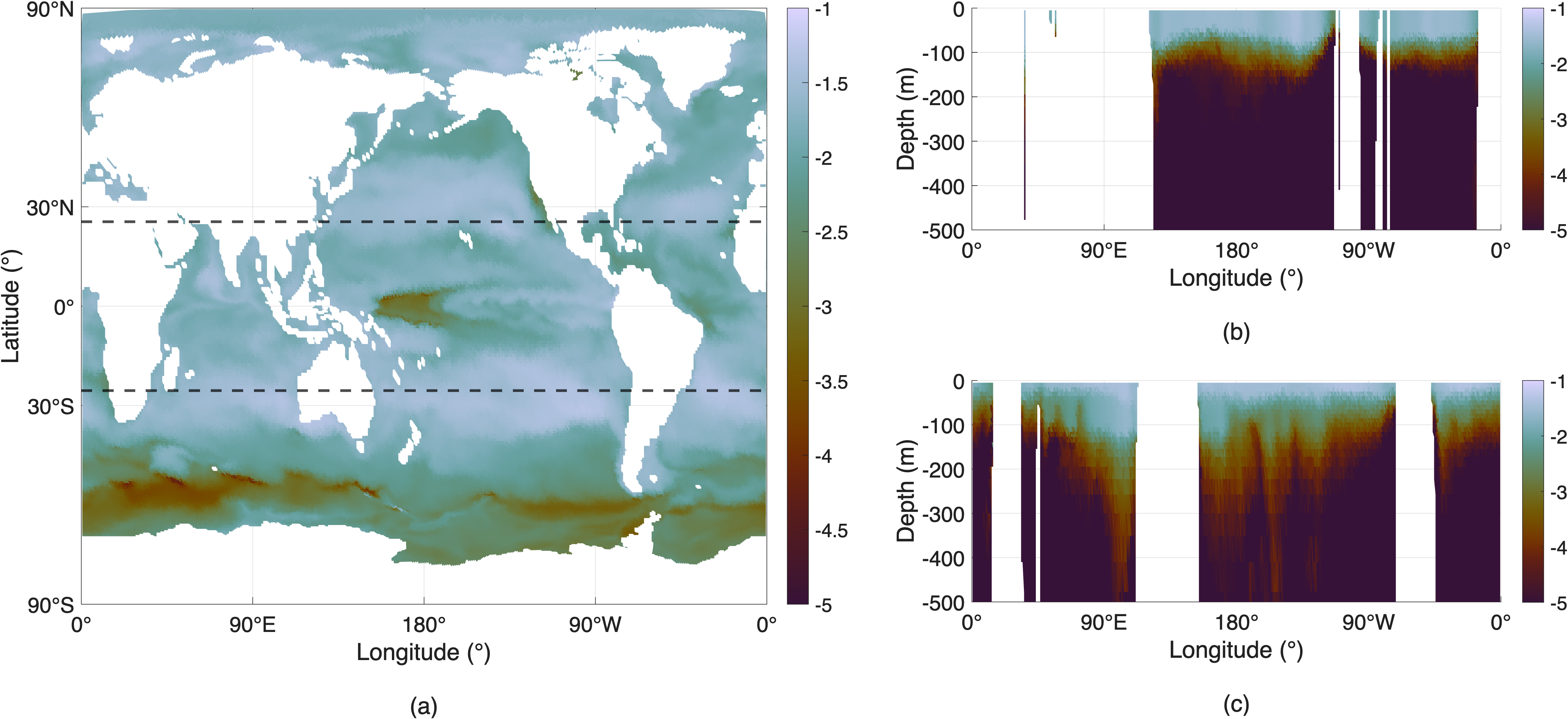}
\caption{(a) The local power-law slope of the spectrum on the sea surface at the end of simulation, with black dashed lines overlaid at $25\degree$N and $25\degree$S. (b) The local power-law slope of the spectrum across $25\degree$N covering the North Pacific gyre. (c) The local power-law slope of the spectrum across $25\degree$N covering the North Atlantic gyre.}
\label{fig:slope_global3d}
\end{figure}

\begin{figure}[!ht]
\centering
\includegraphics[width=0.6\textwidth]{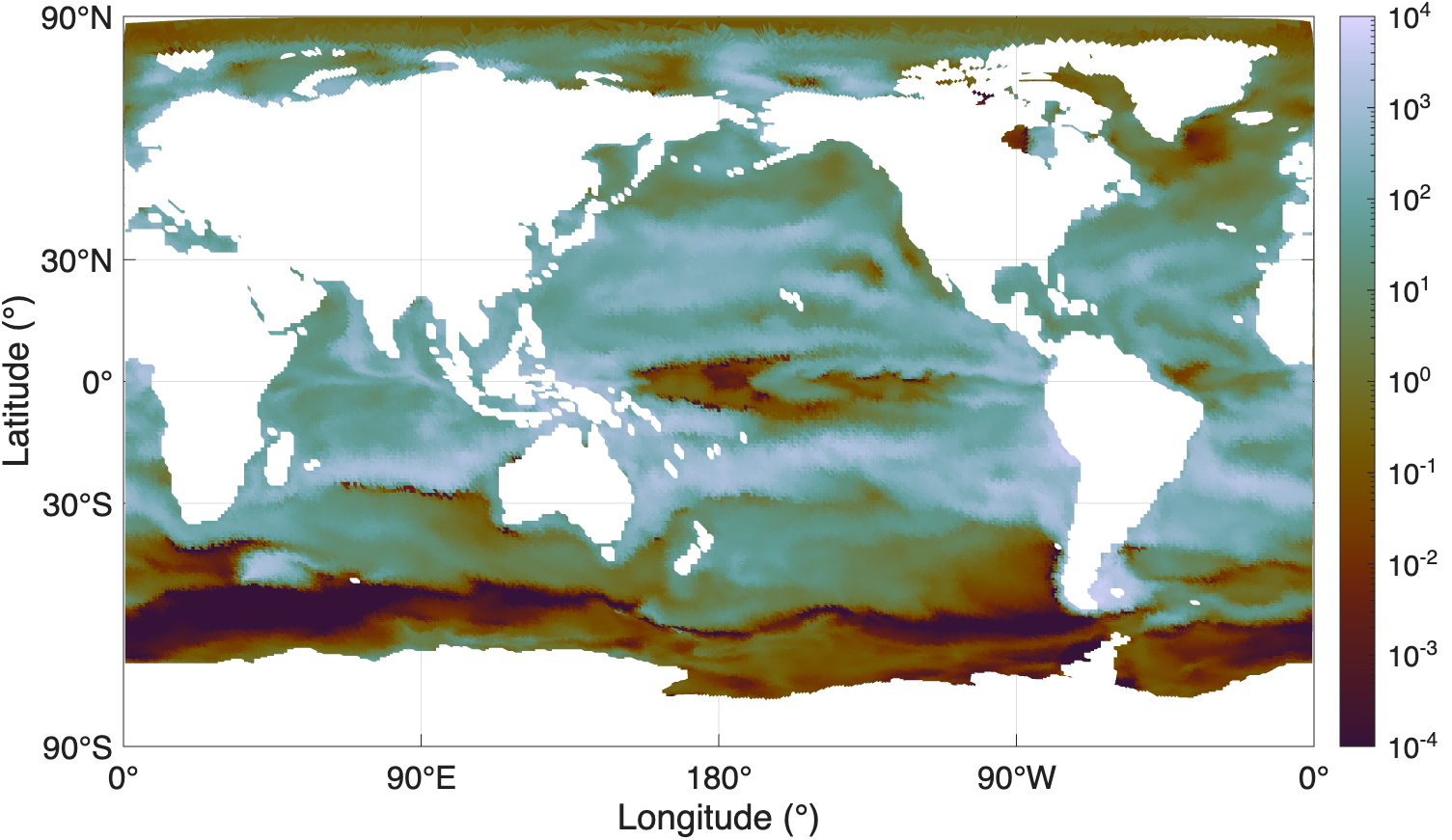}
\caption{The relative concentration of large particles compared to small particles $M_{10}/M_1$ on the sea surface at the end of simulation.}
\label{fig:relative_bin10_bin1}
\end{figure}

The previous section (Sec.~\ref{sec:results_spectrum}) discussed the globally integrated spectrum. Because particles of different sizes follow distinct transport pathways by virtue of their size-dependent terminal velocities, the global spectrum may not be mirrored locally. Here we show the slope of the local spectrum $\tau(\vec{x},m,t)$ (Figure~\ref{fig:slope_global3d}), which varies systematically across the three-dimensional structure of the garbage patches. To facilitate the reader's understanding, we also provide in Figure~\ref{fig:relative_bin10_bin1} the surface map of the ratio $M_{10}/M_1$, which measures the relative mass concentration of the largest to the smallest size bin and serves as a more intuitive proxy for the local spectral slope.

%%%%%%%%%%%%%%
%%%%%%%%%%%%%%

%%%%%%%%%%%%%%
%%%%%%%%%%%%%%

At the sea surface (Figure~\ref{fig:slope_global3d}a), the local slope peaks at the gyre centers with values close to $\sim-1$ and decreases away from them to values of $\sim-4$. This spatial pattern is directly reflected in the relative concentration map (Figure~\ref{fig:relative_bin10_bin1}): $M_{10}/M_1$ is highest at the gyre centers, indicating that large particles dominate there, and falls below $1$ toward the tropical and polar regions, where small particles become relatively more abundant. The steeper slope away from the gyres therefore follows directly from the preferential transport of larger particles into the gyre cores, leaving the surrounding regions enriched in smaller, more neutrally buoyant fragments.

Vertically (Figures~\ref{fig:slope_global3d}b and \ref{fig:slope_global3d}c), the local slope of the spectrum peaks at the sea surface with values close to $\sim-1$ and decreases downward to a value of $<-5$. This behavior is also consistent with the vertical distribution of  particles from different mass bins in Figure~\ref{fig:individuals}. As the particle diameter decreases, their terminal velocity decreases. The weaker terminal velocity provides less buoyancy, and thus smaller particles are more easily transported below the mixed layer. That is, moving downward from the sea surface, fewer large-size particles and more small-size particles are expected to be found. Therefore, the local slope decreases from the sea surface toward the interior ocean.

%%--------------------------------------------------------------------
%%--------------------------------------------------------------------
\subsection{Comparison with observational data}
\label{sec:results_validation}

\begin{figure}[!ht]
\centering
\includegraphics[width=1.0\textwidth]{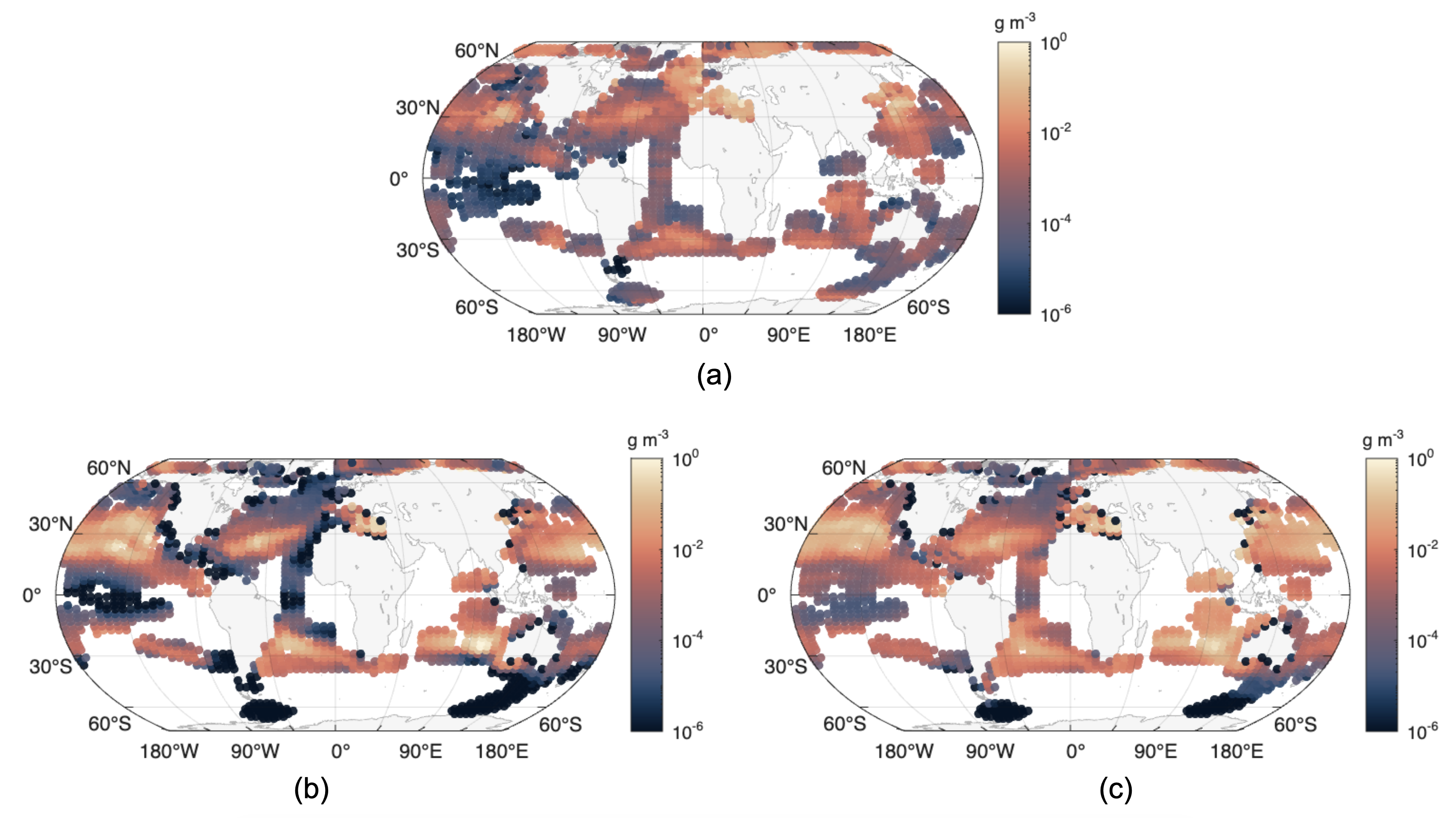}
\caption{(a) The AOMI Level-3wm gridded surface mass concentration field. (b) The modeled fixed-size MP concentration field at the sea surface. (c) The modeled fragmenting MP concentration field at the sea surface.}
\label{fig:validation-surface}
\end{figure}

Having characterized the model's behavior, we compare the simulation results with the AOMI Level-3wm data \citep{isobe_multilevel_2021}, which provides a gridded surface mass concentration field (with a $5\degree$-longitude by $2\degree$-latitude grid spacing) compiled from extensive global trawler observations. The AOMI Level-3wm concentration field is shown in Figure~\ref{fig:validation-surface}a. Some portions of the Indian and South Atlantic oceans are not as thoroughly covered as the North Pacific and North Atlantic oceans by the dataset, because those regions contain fewer sampling points in the raw Level-0 data \citep{isobe_multilevel_2021}. The modeled concentration fields (the same as those shown in Figure~\ref{fig:collective_surface}) are bin-averaged onto the AOMI grid and shown in Figures~\ref{fig:validation-surface}b and \ref{fig:validation-surface}c. 

%%%%%%%%%%%%%%
%%%%%%%%%%%%%%

The modeled size range (0--50~$\mu\text{m}$) and the size range resolved by the AOMI dataset (0.3--5~mm) do not overlap. Thus, instead of seeking value-matching, our comparison targets spatial pattern correlation between the observed and the modeled concentration fields. The spatial correlation between surface concentrations is evaluated using the area-weighted Pearson correlation coefficient. Since the area of each cell on a regular grid varies with the cosine of the latitude, $\cos\phi_i$, each observation point~$i$ is assigned a weight $w_i = \cos\phi_i$. The weighted Pearson correlation coefficient between the log-transformed observed concentration $x_i = \log_{10} M_i^\text{obs}$ and modeled concentration $y_i = \log_{10} M_i^\text{mod}$ is defined as
%%%%%%%%%%%%%%
\begin{equation}
    R_w = \frac{\displaystyle\sum_i w_i \,(x_i - \bar{x}_w)\,(y_i - \bar{y}_w)}
               {\sqrt{\displaystyle\sum_i w_i\,(x_i - \bar{x}_w)^2 \;\cdot\; 
                      \displaystyle\sum_i w_i\,(y_i - \bar{y}_w)^2}},
    \label{eq:weighted_corr}
\end{equation}
%%%%%%%%%%%%%%
where $\bar{x}_w = \sum_i w_i x_i \,/\, \sum_i w_i$ and $\bar{y}_w = \sum_i w_i y_i \,/\, \sum_i w_i$ are the weighted means. The spatial correlation between the AOMI Level-3wm observations and the modeled surface concentrations improves from 45\% for the fixed-size simulation to 58\% when fragmentation is included.

%%%%%%%%%%%%%%
\begin{figure}[!ht]
\centering
\includegraphics[width=1.0\textwidth]{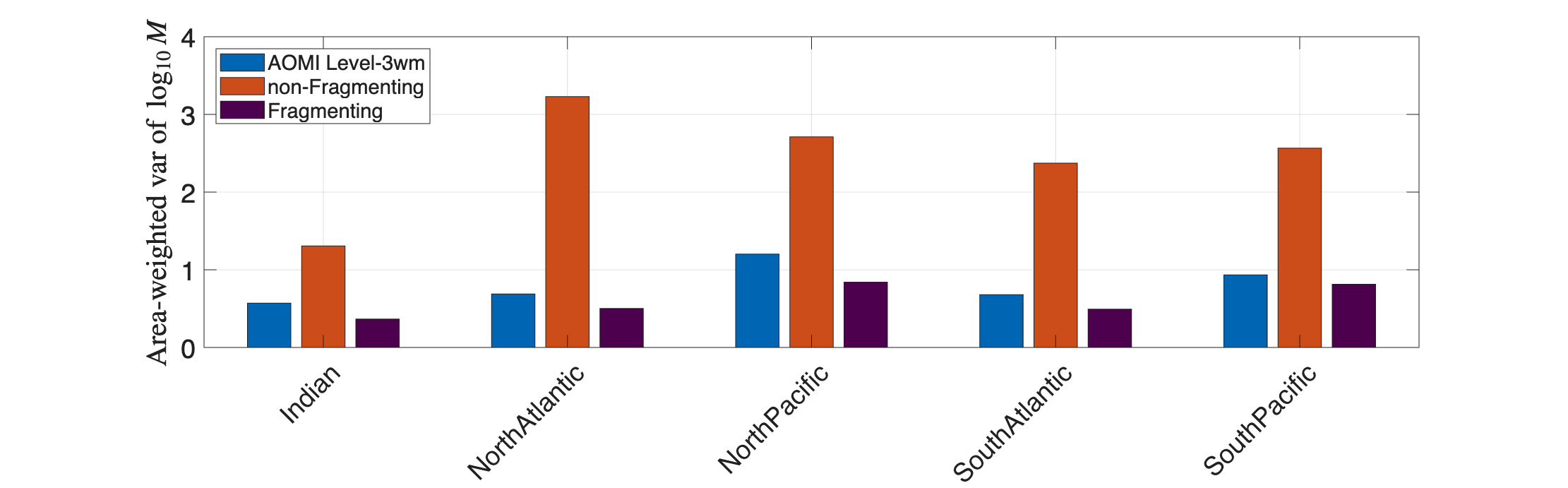}
\caption{The observed (AOMI Level-3wm) vs. modeled (with and without fragmentation) variance of surface mass concentration $M$.}
\label{fig:validation-variance}
\end{figure}
%%%%%%%%%%%%%%
To quantify the surface-smoothing effect of fragmentation, we evaluate the variance of the concentration field within each ocean basin, which measures the contrast between high- and low-concentration regions \citep{tseng_effect_2026}. Using the same weights $w_i = \cos\phi_i$ and weighted mean $\bar{x}_w$ defined in Eq.~\eqref{eq:weighted_corr}, the area-weighted variance within a basin $\mathcal{B}$ is
%%%%%%%%%%%%%%
\begin{equation}
    \sigma^2_w = \frac{\displaystyle\sum_{i\in\mathcal{B}} w_i\,(x_i - \bar{x}_w)^2}
                      {\displaystyle\sum_{i\in\mathcal{B}} w_i},
    \label{eq:weighted_var}
\end{equation}
%%%%%%%%%%%%%%
where $x_i = \log_{10}M_i$ is the log-transformed concentration at point $i$ and the sum runs over all valid observation points within $\mathcal{B}$. As shown in Figure~\ref{fig:validation-variance}, incorporating fragmentation reduces $\sigma^2_w$ toward the observed values across all major ocean basins, consistent with the horizontal smoothing identified in Sec.~\ref{sec:results_collective}.

%%%%%%%%%%%%%%%%%%%%%%%%%%%%%%%%%%%%%%%%%%%%%%%%%%%%%%%%%%%%%%%%%%%%%
%%%%%%%%%%%%%%%%%%%%%%%%%%%%%%%%%%%%%%%%%%%%%%%%%%%%%%%%%%%%%%%%%%%%%
\section{Conclusion}
\label{sec:conclusion}

This study presents, to our knowledge, the first global 3D Eulerian ocean model that explicitly resolves the fragmentation of MPs alongside their transport. By coupling the fragmentation kinetics with the advection-diffusion equation, the model captures both the continuous evolution of particle size and the resulting redistribution of MPs in the ocean, which prior models cannot capture.

Comparing between the fragmenting-particles and the fixed-size-particles simulations reveals some noteworthy differences. In terms of the collective mass concentration, fragmentation results in more horizontally dispersed surface garbage patches in the five subtropical gyres and a vertical downshifting to roughly $500$~m depth. This is because fragmentation continuously produces smaller particles with weaker buoyancy, which are more easily entrained below the mixed layer and are subject to weaker Ekman transport toward gyre centers.

The collective concentration is a superposition of distinct mass-dependent patterns, and we examine the contribution of particles from individual mass bins. The largest particles, which dominate the total mass, remain confined to the mixed layer and reproduce the well-known subtropical garbage patches. In contrast, smaller particles can sink below the mixed layer and are transported along subsurface pathways. In particular, neutrally buoyant particles exhibit peak concentrations at the centers of the subtropical gyres because they are generated in situ through the fragmentation of larger particles that have already accumulated there. The improved representation of this in situ source of small particles represents a step forward from earlier studies that did not consider fragmentation, which could bias the simulated distribution by overestimating concentrations near coastlines and even eliminating the surface garbage patches in the five subtropical gyres.

The evolution of the mass density spectrum of particle number concentration provides more details from both global and local points of view. Globally, the integrated spectrum maintains a power-law form throughout the simulation, consistent with linear fragmentation theory \citep{cozar_plastic_2014,sorasan_ageing_2022,cheng_kinetics_1990,ben_naim_fragmentation_2000}. The slope of the spectrum continues to steepen and has not reached the theoretical steady state after 25 years of simulation time. Locally, the slope of the spectrum decreases both horizontally away from the gyre centers and vertically downward from the sea surface. This pattern is consistent with the mass-dependent transport: large particles remain near the surface and within the gyres, while small particles preferentially spread to depth and to the periphery.

Finally, comparison with the AOMI Level-3wm dataset confirms that accounting for fragmentation dynamics yields a meaningful improvement in model skill. Including fragmentation raises the spatial correlation between modeled and observed surface concentrations from 45\% to 58\%; the variance of mass concentration fields is reduced by~7 times in the North Atlantic ocean and now matches observations better across all major ocean basins. Together, these results show that fragmentation is an important process that governs where and how microplastics are generated and destined in the global ocean.

Several limitations of the current study highlight key directions for future research. First, fragmentation-generated microplastics are subject to biological and ecological processes that act as material sinks, such as trophic transfer within marine food webs or settling into benthic ecosystems. Currently, these removal pathways are not resolved by our ocean transport model due to a lack of available parameterizations. Second, the particles considered in this study are restricted to the 0--50~$\mu\text{m}$ range, deliberately small enough that the underlying Stokes-flow assumption of Eq.~\eqref{eq:terminal_velocity} remains valid. Extending the model to resolve larger MPs and macroplastics would require: (1) a terminal velocity formula that applies beyond the Stokes regime, where particle geometry comes into play through inertial effects, and (2) a substantially smaller simulation timestep to resolve the correspondingly faster vertical transport of more buoyant particles, which adds considerable computational cost. Third, the present model resolves fragmentation alone and does not consider bio-interactions \citep{richon_faecal_2026}, which we have shown in a companion study \citep{tseng_effect_2026} to alter the distribution of MPs through the mass added by the attaching biofilm. Because fragmentation and biofouling both act on particle buoyancy---one by reducing particle size and the other by increasing effective density---their combined effect on the global MP distribution remains unresolved, and likely involves even higher dimensions of parameters (particle diameter $d$ and algal attachment $A$, among others).

Given these limitations, the present study focuses on establishing the first-order importance of fragmentation in reshaping the global distribution of MPs across a size range for which the modeling assumptions remain robust. Building on this foundation, our next step is to incorporate data assimilation into this modeling framework. Assimilating observations from growing MP datasets, such as the CYGNSS Level-3 Ocean Microplastic Concentration Version 3.2 product \citep{cygnss_team_cygnss_2024}, into our model should further improve the predictive skill of the simulation.

%%%%%%%%%%%%%%%%%%%%%%%%%%%%%%%%%%%%%%%%%%%%%%%%%%%%%%%%%%%%%%%%%%%%%
\section*{Data Availability Statement}

To reproduce simulations in this paper, one can follow five procedures briefly summarized below.
\begin{enumerate}
   \item Download MITgcm source code.
   \item Download ECCO V4r5 dataset (in particular the forcing and initial conditions for MITgcm to reproduce the ECCO V4r5 dataset).
   \item Download additional code (used in step 4) and inputs (used in step 5) at \url{https://zenodo.org/records/21537418} (recommended) or \url{https://github.com/zizien1019/fragmentation_eccov4r5_mitgcm68o.git} to treat additional terms in Eq.~\eqref{eq:adv-diff}. Example outputs of the simulation are also provided in the repository.
   \item Compile both the original MITgcm code and additional code in step 3.
   \item Conduct simulations with different inputs on particle properties.
\end{enumerate}

The README file associated with the GitHub repository \url{https://github.com/zizien1019/fragmentation_eccov4r5_mitgcm68o.git} contains much more detailed step-by-step instructions.

%%%%%%%%%%%%%%%%%%%%%%%%%%%%%%%%%%%%%%%%%%%%%%%%%%%%%%%%%%%%%%%%%%%%%
\section*{Acknowledgments}

This research was supported in part by NASA Science Mission Directorate contract 80LARC21DA003 with the University of Michigan.

This work used Bridges-2 \citep{psc_bridges2_2024} at the Pittsburgh Supercomputing Center through allocations phy250044p and ees230007p from the ACCESS program \citep{boerner_access_2023}, which is supported by U.S. National Science Foundation grants \#2138259, \#2138286, \#2138307, \#2137603, and \#2138296.

Portions of this research were conducted with the advanced computing resources provided by Texas A\&M High Performance Research Computing.

%%%%%%%%%%%%%%%%%%%%%%%%%%%%%%%%%%%%%%%%%%%%%%%%%%%%%%%%%%%%%%%%%%%%%
%%%%%%%%%%%%%%%%%%%%%%%%%%%%%%%%%%%%%%%%%%%%%%%%%%%%%%%%%%%%%%%%%%%%%
\appendix

%%%%%%%%%%%%%%%%%%%%%%%%%%%%%%%%%%%%%%%%%%%%%%%%%%%%%%%%%%%%%%%%%%%%%
%%%%%%%%%%%%%%%%%%%%%%%%%%%%%%%%%%%%%%%%%%%%%%%%%%%%%%%%%%%%%%%%%%%%%
\section{Steady-state solution to the fragmentation equation}
\label{append:steady-state}

%%--------------------------------------------------------------------
%%--------------------------------------------------------------------
\subsection{Analytical form}
\label{append:steady-state-analytical}

For Eq.~\eqref{eq:fragmentation}, we seek the steady-state solution $\tau(m) \equiv \lim_{t\to\infty} \tau(m,t)$ for $m < m_\text{max}$. At steady state, the time derivative vanishes and the governing equation becomes
%%%%%%%%%%%%%%
\begin{equation}
0 = -a(m) \,\tau(m) + \int_{m}^{m_\text{max}} a(\eta) \,b(m|\eta) \,\tau(\eta) \,d\eta,
\label{eq:steady-state}
\end{equation}
%%%%%%%%%%%%%%
with the boundary condition $\tau(m_\text{max}) = c_L$. We assume the solution takes a power-law form $\tau(m) = C m^n$, and substitute $a(m) = a_0 m^\lambda$ and $b(m|\eta) = \frac{\nu+2}{\eta}\left(\frac{m}{\eta}\right)^\nu$ into Eq.~\eqref{eq:steady-state}, resulting in
%%%%%%%%%%%%%%
\begin{equation}
0 = -a_0 C\, m^{n+\lambda} + a_0 C (\nu+2) m^\nu \int_m^{m_\text{max}} \eta^{\lambda-\nu+n-1} \, d\eta.
\end{equation}
%%%%%%%%%%%%%%

Evaluating the integral gives
%%%%%%%%%%%%%%
\begin{equation}
0 = -a_0 C\, m^{n+\lambda} + a_0 C (\nu+2) m^\nu \cdot \frac{m_\text{max}^{\lambda-\nu+n} - m^{\lambda-\nu+n}}{\lambda-\nu+n},
\end{equation}
%%%%%%%%%%%%%%
which holds for all $m \ll m_\text{max}$, thus the $m_\text{max}$-dependent term is negligible and the remaining two terms must balance. Collecting the $m^{n+\lambda}$ terms gives
%%%%%%%%%%%%%%
\begin{equation}
0 = -1 - \frac{\nu+2}{\lambda-\nu+n},
\end{equation}
%%%%%%%%%%%%%%
which gives $n = -(\lambda+2)$. Applying the boundary condition $\tau(m_\text{max}) = c_L$ fixes the prefactor $C = c_L \,m_\text{max}^{\lambda+2}$, yielding the solution Eq.~\eqref{eq:steady-state-solution}: $\lim_{t\to\infty} \tau(m,t) = c_L \,\left(\dfrac{m}{m_\text{max}}\right)^{-(\lambda+2)}$.

%%--------------------------------------------------------------------
%%--------------------------------------------------------------------
\subsection{Numerical confirmation}
\label{append:steady-state-numerical}

To verify the analytical results derived in the preceding subsection (\ref{append:steady-state-analytical}) and to characterize the timescale required to reach steady state, we numerically integrate the fragmentation equations over long durations across multiple combinations of the parameters $\lambda$ and $\nu$. For each combination, we record the resulting steady-state power-law slope and the time required to achieve it.

Table~\ref{tab:steadystate_params} summarizes these results. Two features are evident. First, for fixed $\nu$, increasing $\lambda$ increases the time required to reach steady state. Second, the steady-state slope depends only on $\lambda$ and is independent of $\nu$, i.e., for a given $\lambda$, the slope is identical across all three values of $\nu$ tested. This confirms the analytical prediction $n=-(\lambda+2)$ derived in \ref{append:steady-state-analytical}, in which $\nu$ does not appear.

%%%%%%%%%%%%%%
\begin{table}[!ht]
\centering
\caption{Steady-state power-law slope and the time required to reach it (in years, in the parentheses) for combinations of $\lambda$ and $\nu$. The steady-state slope depends only on $\lambda$, consistent with the analytical solution $n=-(\lambda+2)$; the time required to reach steady state depends on both parameters and is on the order of several centuries.}
\label{tab:steadystate_params}
\begin{tabular}{c ccc}
\toprule
\multirow{2}{*}{$\nu$} & \multicolumn{3}{c}{$\lambda$} \\
\cmidrule(lr){2-4}
 & $0.01$ & $0.02$ & $0.05$ \\
\midrule
$-1.5$   & $-2.01$ (375 yrs) & $-2.02$ (375 yrs) & $-2.05$ (750 yrs) \\
$-1.667$ & $-2.01$ (250 yrs) & $-2.02$ (375 yrs) & $-2.05$ (625 yrs) \\
$-1.8$   & $-2.01$ (250 yrs) & $-2.02$ (250 yrs) & $-2.05$ (500 yrs) \\
\bottomrule
\end{tabular}
\end{table}
%%%%%%%%%%%%%%

%%%%%%%%%%%%%%%%%%%%%%%%%%%%%%%%%%%%%%%%%%%%%%%%%%%%%%%%%%%%%%%%%%%%%
%%%%%%%%%%%%%%%%%%%%%%%%%%%%%%%%%%%%%%%%%%%%%%%%%%%%%%%%%%%%%%%%%%%%%
\section{Discretization on the dimension of $m$}
\label{append:discretization}

Here we derive the discrete form of the fragmentation equation, Eq.~\eqref{eq:fragmentation}, where the mass conversion rate is the central quantity relating its continuous and discrete forms. 

We start with defining the mass bins and some properties related to the lower- and upper-bound masses of a bin. The dimension of particle mass is divided into $N=10$ bins, with the left end, right end, and center points of the $i^{th}$ bin defined to be $L_i$, $R_i$, and $m_i=(R_i+L_i)/2$, respectively. In the formulation of Eq.~\eqref{eq:fragmentation}, the spectrum $\tau(m,t)$ (in unit \# m$^{-3}$ g$^{-1}$) is continuous on $m$. While in the simulation, the working quantity is the number concentration $c_i \equiv c_i (m_i,t)$ allocated at bin centers and representing the collection of all particles with mass in the range $m \in (L_i, R_i]$
%%%%%%%%%%%%%%
\begin{equation}
    c_i = \int^{R_i}_{L_i} \tau(m,t) \, dm.
\end{equation}
%%%%%%%%%%%%%%
We then take the average concentration within each interval $(L_i, R_i]$ to construct a piecewise constant spectrum $\tau_{pc}(m,t)$ to approximate the continuous spectrum $\tau(m,t)$
%%%%%%%%%%%%%%
\begin{equation}
    \tau_{pc}(m,t) = \frac{c_i}{R_i-L_i} = \frac{1}{R_i-L_i} \int^{R_i}_{L_i} \tau(m,t) \,dm, \quad \mbox{for} \quad L_i < m \leq R_i.
\end{equation}
%%%%%%%%%%%%%%
The collective mass of particles with $m \in (L_i, R_i]$ then equals
%%%%%%%%%%%%%%
\begin{equation}
    M_i(t) = \int^{R_i}_{L_i} m \,\tau_{pc}(m,t) \, dm = \int^{R_i}_{L_i} m \, \frac{c_i}{R_i-L_i}  \, dm = \frac{R_i+L_i}{2} c_i = m_i \, c_i.
    \label{eq:frag-mass-conserv}
\end{equation}
%%%%%%%%%%%%%%

Substituting $\tau_{pc}(m,t)$ into the continuous fragmentation equation, Eq.~\eqref{eq:fragmentation} gives the evolution of the piecewise constant spectrum
%%%%%%%%%%%%%%
\begin{equation}
    \frac{d\tau_{pc}(m,t)}{dt} = -a(m) \, \tau_{pc}(m,t) + \int_{m}^{R_N} a(\eta) \, b(m|\eta) \, \tau_{pc}(\eta,t) \, d\eta + Q\,\delta(m-m_\text{max}).
    \label{eq:frag-piece-wise}
\end{equation}
%%%%%%%%%%%%%%

To obtain an evolution equation for $c_i$, we integrate Eq.~\eqref{eq:frag-piece-wise} over $m$. This converts the reduction and gain terms on the right-hand side of Eq.~\eqref{eq:frag-piece-wise} into a matrix operation acting on the vector of binned concentrations $c_j$. The resulting discrete fragmentation system reads
%%%%%%%%%%%%%%
\begin{align}
    \frac{\partial c_i}{\partial t} 
        & = - A_{i} \, c_i + \sum_{j=i}^N  B_{ij} \, c_j + Q\,\delta_{iN}, \\
        & = \sum_{j=i}^N (- A_{i} \, \delta_{ij} + B_{ij}) \, c_j + Q\,\delta_{iN},
    \label{eq:frag_discrete}
\end{align}
%%%%%%%%%%%%%%
with $\delta_{ij}$ being the Kronecker delta function. The right-hand side is expressed through (1) a diagonal matrix $-A_i\,\delta_{ij}$, capturing the loss of mass leaving bin $i$; and (2) an upper-triangular matrix $B_{ij}$, capturing the gain of mass entering bin $i$ from every larger bin $j\geq i$. The conservation of mass, Eq.~\eqref{eq:frag-mass-conserv}, provides the constraint through which Eq.~\eqref{eq:frag-piece-wise} and Eq.~\eqref{eq:frag_discrete} are related, and from which the explicit forms of $A_i$ and $B_{ij}$ are obtained:
%%%%%%%%%%%%%%
\begin{equation}
    A_i = \frac{2a_0}{\lambda+2} \, \frac{R_i^{\lambda+2} - L_i^{\lambda+2}}{R_i^2 - L_i^2},
    \label{eq_frag_Ai_preview}
\end{equation}
%%%%%%%%%%%%%%
\begin{align}
    B_{ij} = 
    \begin{cases}
        \dfrac{R_i - L_i}{R_j - L_j} \, \dfrac{2a_0}{R_i^2 - L_i^2} \, \dfrac{R_j^{\lambda-\nu} - L_j^{\lambda-\nu}}{\lambda - \nu} \, (R_i^{\nu+2} - L_i^{\nu+2}),
        & i < j, \\[3ex]
        \dfrac{2a_0}{R_i^2 - L_i^2} \left[ \dfrac{R_i^{\lambda+2} - L_i^{\lambda+2}}{\lambda+2} 
        - L_i^{\nu+2} \, \dfrac{R_i^{\lambda-\nu} - L_i^{\lambda-\nu}}{\lambda-\nu} \right], 
        & i = j.
    \end{cases}
    \label{eq_frag_Bij_preview}
\end{align}
%%%%%%%%%%%%%%

The following two sections~\ref{append:A_i} and \ref{append:B_ij} present this derivation for $A_i$ and $B_{ij}$ in turn, with mass conservation enforced throughout, so that the mass conversion rate due to the discrete exchange is consistent with the continuous fragmentation equation.

%%--------------------------------------------------------------------
%%--------------------------------------------------------------------
\subsection{Derivation of $A_i$ (reduction rate)}
\label{append:A_i}

The reduction term represents particles leaving bin $i$ due to fragmentation. Examining the loss of mass within each bin gives
%%%%%%%%%%%%%%
\begin{equation}
    \int^{R_i}_{L_i} m \, \left[ -a(m) \, \tau_{pc}(m) \right] \, dm =  \int^{R_i}_{L_i} m \, \left[ -A_i\,\frac{c_i}{R_i-L_i} \right] \, dm.
\end{equation}
%%%%%%%%%%%%%%
Note that within bin $i$, $\tau_{pc}(m) = \dfrac{c_i}{R_i - L_i}$ is constant, and we have
%%%%%%%%%%%%%%
\begin{align}
    A_i &= \frac{\displaystyle\int^{R_i}_{L_i} m \, a(m) \, \tau_{pc}(m) \, dm}{\displaystyle\int^{R_i}_{L_i} m \, \frac{c_i}{R_i-L_i} \, dm} \nonumber \\
        &= \frac{\displaystyle\int^{R_i}_{L_i} m \, a(m) \, dm}{\displaystyle\int^{R_i}_{L_i} m \, dm}.
        \label{eq_frag_Ai}
\end{align}
%%%%%%%%%%%%%%
Substituting $a(x)=a_0\, x^\lambda$ into Eq.~\eqref{eq_frag_Ai} gives
%%%%%%%%%%%%%%
\begin{align}
    A_i &= \frac{2a_0}{\lambda+2} \, \frac{R_i^{\lambda+2} - L_i^{\lambda+2}}{R_i^2 - L_i^2}.
    \label{eq_frag_Ai_spellout}
\end{align}
%%%%%%%%%%%%%%

%%--------------------------------------------------------------------
%%--------------------------------------------------------------------
\subsection{Derivation of $B_{ij}$ (gain rate)}
\label{append:B_ij}

Following the same approach as in \ref{append:A_i}, we examine the mass gained within each bin $i$, which equals
%%%%%%%%%%%%%%
\begin{equation}
    \int^{R_i}_{L_i} m \, \left[ \int_{m}^{R_N} a(\eta) \, b(m|\eta) \, \tau_{pc}(\eta) \, d\eta \right] \, dm.
    \label{eq:gain-original}
\end{equation}
%%%%%%%%%%%%%%
For a given $m \in [L_i, R_i]$, the inner integral on $\eta \in [m, R_N]$ can be split into the contributions from bin $i$ itself on $\eta \in [m, R_i]$, and from every larger-mass bin $j>i$ on $\eta \in [L_{i+1}, R_N]$
%%%%%%%%%%%%%%
\begin{equation}
    \int_m^{R_N} a(\eta)\,b(m|\eta)\,\tau_{pc}(\eta)\,d\eta = \int_m^{R_i} a(\eta)\,b(m|\eta)\,\tau_{pc}(\eta)\,d\eta + \sum_{j=i+1}^N \int_{L_j}^{R_j} a(\eta)\,b(m|\eta)\,\tau_{pc}(\eta)\,d\eta.
    \label{eq:gain-split}
\end{equation}
%%%%%%%%%%%%%%
Substituting Eq.~\eqref{eq:gain-split} into Eq.~\eqref{eq:gain-original} decomposes the gain term within bin $i$ into a within-bin contribution and a sum of cross-bin contributions from every larger-mass bin
%%%%%%%%%%%%%%
\begin{equation}
    \eqref{eq:gain-original} = \underbrace{\int^{R_i}_{L_i} m \left[ \int_m^{R_i} a(\eta)\,b(m|\eta)\,\tau_{pc}(\eta)\,d\eta \right] dm}_{\text{within-bin, } j=i} + \sum_{j=i+1}^N \underbrace{\int^{R_i}_{L_i} m \left[ \int_{L_j}^{R_j} a(\eta)\,b(m|\eta)\,\tau_{pc}(\eta)\,d\eta \right] dm}_{\text{cross-bin, } j>i}.
    \label{eq:gain-decomposed}
\end{equation}
%%%%%%%%%%%%%%

This piecewise constant form of the gain-induced mass conversion rate, Eq.~\eqref{eq:gain-decomposed}, must equal the corresponding discrete form from Eq.~\eqref{eq:frag_discrete}
%%%%%%%%%%%%%%
\begin{equation}
    \eqref{eq:gain-decomposed} = \int^{R_i}_{L_i} m \, \frac{1}{R_i-L_i} \left[ \sum_{j=i}^N B_{ij}\,c_j \right] dm = \sum_{j=i}^N B_{ij}\,\frac{c_j}{R_i-L_i}\int_{L_i}^{R_i} m\,dm.
    \label{eq:gain-discrete-form}
\end{equation}
%%%%%%%%%%%%%%
Equating Eq.~\eqref{eq:gain-decomposed} and Eq.~\eqref{eq:gain-discrete-form} term-by-term---the within-bin term ($j=i$) on the left with the $j=i$ term on the right, and each cross-bin term ($j>i$) on the left with the corresponding $j>i$ term on the right---identifies $B_{ij}$ for each pair $(i,j)$ with $i\leq j$
%%%%%%%%%%%%%%
\begin{align}
    B_{ij} = 
    \begin{cases}
        \dfrac{\displaystyle\int^{R_i}_{L_i} m \left[ \displaystyle\int_m^{R_i} a(\eta)\,b(m|\eta)\,\tau_{pc}(\eta)\,d\eta \right] dm }{\displaystyle\dfrac{c_i}{R_i-L_i} \int_{L_i}^{R_i} m \, dm}, & \mbox{if} \quad i = j, \\[3ex]
        \dfrac{\displaystyle\int^{R_i}_{L_i} m \left[ \displaystyle\int_{L_j}^{R_j} a(\eta)\,b(m|\eta)\,\tau_{pc}(\eta)\,d\eta \right] dm}{\displaystyle\dfrac{c_j}{R_i-L_i}\int_{L_i}^{R_i} m \, dm}, & \mbox{if} \quad i < j.
    \end{cases}
    \label{eq_frag_Bij_abstract}
\end{align}
%%%%%%%%%%%%%%
The remainder of this section evaluates the two integral expressions in Eq.~\eqref{eq_frag_Bij_abstract} explicitly, treating the within-bin ($j=i$) and cross-bin ($j>i$) cases in turn.

\textbf{Within-bin contribution ($j=i$).} Here both $m$ and $\eta$ lie within the same bin $[L_i,R_i]$, with the constraint $m\leq\eta\leq R_i$
%%%%%%%%%%%%%%
\begin{align}
    \int^{R_i}_{L_i} m \left[ \int_m^{R_i} a(\eta)\,b(m|\eta)\,\tau_{pc}(\eta)\,d\eta \right] dm 
    &= \frac{c_i}{R_i-L_i} \int^{R_i}_{L_i} m \left[ \int_{m}^{R_i} a(\eta) \, b(m|\eta) \, d\eta \right] dm.
    \label{eq:gain-withinbin-prefactored}
\end{align}
%%%%%%%%%%%%%%
Comparing Eq.~\eqref{eq:gain-withinbin-prefactored} with the $i=j$ case of Eq.~\eqref{eq_frag_Bij_abstract} gives
%%%%%%%%%%%%%%
\begin{equation}
    B_{ii} = \frac{\displaystyle\int^{R_i}_{L_i} m \left[ \displaystyle\int_{m}^{R_i} a(\eta)\,b(m|\eta)\,d\eta \right] dm}{\displaystyle\int_{L_i}^{R_i} m\,dm}.
    \label{eq:Bii-preform}
\end{equation}
%%%%%%%%%%%%%%
Substituting $a(\eta)=a_0\,\eta^\lambda$ and $b(m|\eta)=\dfrac{\nu+2}{\eta}\left(\dfrac{m}{\eta}\right)^{\!\nu}$ into the inner integral of Eq.~\eqref{eq:Bii-preform} gives
%%%%%%%%%%%%%%
\begin{align}
    \int_{m}^{R_i} a(\eta)\,b(m|\eta)\,d\eta
    &= a_0(\nu+2)\,m^\nu \int_{m}^{R_i} \eta^{\lambda-\nu-1}\,d\eta \nonumber \\
    &= \frac{a_0(\nu+2)}{\lambda-\nu}\,m^\nu\left(R_i^{\lambda-\nu}-m^{\lambda-\nu}\right).
    \label{eq:Bii-eta-integral}
\end{align}
%%%%%%%%%%%%%%
Substituting Eq.~\eqref{eq:Bii-eta-integral} into the outer integral of Eq.~\eqref{eq:Bii-preform} then gives
%%%%%%%%%%%%%%
\begin{align}
    \int^{R_i}_{L_i} m \left[ \int_{m}^{R_i} a(\eta)\,b(m|\eta)\,d\eta \right] dm
    &= \frac{a_0(\nu+2)}{\lambda-\nu} \int_{L_i}^{R_i} \left[ R_i^{\lambda-\nu}\,m^{\nu+1} - m^{\lambda+1} \right] dm \nonumber \\
    &= a_0\left[\frac{R_i^{\lambda+2}-L_i^{\lambda+2}}{\lambda+2} - L_i^{\nu+2}\,\frac{R_i^{\lambda-\nu}-L_i^{\lambda-\nu}}{\lambda-\nu}\right].
    \label{eq:Bii-m-integral}
\end{align}
%%%%%%%%%%%%%%
Substituting Eq.~\eqref{eq:Bii-m-integral}, together with $\displaystyle\int_{L_i}^{R_i} m\,dm = (R_i^2-L_i^2)/2$, into Eq.~\eqref{eq:Bii-preform} yields the explicit form
%%%%%%%%%%%%%%
\begin{equation}
    B_{ii} = \frac{2a_0}{R_i^2-L_i^2}\left[\frac{R_i^{\lambda+2}-L_i^{\lambda+2}}{\lambda+2} - L_i^{\nu+2}\,\frac{R_i^{\lambda-\nu}-L_i^{\lambda-\nu}}{\lambda-\nu}\right].
    \label{eq:Bii-final}
\end{equation}
%%%%%%%%%%%%%%

\textbf{Cross-bin contribution ($j>i$).} In this term, $\eta$ ranges over the fixed interval $[L_j, R_j]$ and is independent of $m\in[L_i,R_i]$. Therefore, $\tau_{pc}(\eta) = c_j/(R_j-L_j)$ is a constant within the integration domain and can be factored out
%%%%%%%%%%%%%%
\begin{align}
    \int^{R_i}_{L_i} m \left[ \int_{L_j}^{R_j} a(\eta)\,b(m|\eta)\,\tau_{pc}(\eta)\,d\eta \right] dm 
    &= \frac{c_j}{R_j-L_j} \int^{R_i}_{L_i} m \left[ \int_{L_j}^{R_j} a(\eta)\,b(m|\eta)\,d\eta \right] dm.
    \label{eq:gain-crossbin-prefactored}
\end{align}
%%%%%%%%%%%%%%
Comparing Eq.~\eqref{eq:gain-crossbin-prefactored} with the $i<j$ case of Eq.~\eqref{eq_frag_Bij_abstract} gives
%%%%%%%%%%%%%%
\begin{equation}
    B_{ij} = \frac{R_i-L_i}{R_j-L_j} \, \frac{\displaystyle\int^{R_i}_{L_i} m \left[ \int_{L_j}^{R_j} a(\eta)\,b(m|\eta)\,d\eta \right] dm}{\displaystyle\int_{L_i}^{R_i} m\,dm}, \qquad i<j.
    \label{eq:Bij-crossbin-preform}
\end{equation}
%%%%%%%%%%%%%%
Substituting $a(\eta)=a_0\,\eta^\lambda$ and $b(m|\eta)=\dfrac{\nu+2}{\eta}\left(\dfrac{m}{\eta}\right)^{\!\nu}$ into the inner integral of Eq.~\eqref{eq:Bij-crossbin-preform} gives
%%%%%%%%%%%%%%
\begin{align}
    \int_{L_j}^{R_j} a(\eta)\,b(m|\eta)\,d\eta
    &= a_0(\nu+2)\,m^\nu \int_{L_j}^{R_j} \eta^{\lambda-\nu-1}\,d\eta \nonumber \\
    &= \frac{a_0(\nu+2)}{\lambda-\nu}\,m^\nu\left(R_j^{\lambda-\nu}-L_j^{\lambda-\nu}\right).
    \label{eq:Bij-crossbin-eta-integral}
\end{align}
%%%%%%%%%%%%%%
Substituting Eq.~\eqref{eq:Bij-crossbin-eta-integral} into the outer integral of Eq.~\eqref{eq:Bij-crossbin-preform} then gives
%%%%%%%%%%%%%%
\begin{align}
    \int^{R_i}_{L_i} m \left[ \int_{L_j}^{R_j} a(\eta)\,b(m|\eta)\,d\eta \right] dm
    &= \frac{a_0(\nu+2)}{\lambda-\nu}\left(R_j^{\lambda-\nu}-L_j^{\lambda-\nu}\right) \int_{L_i}^{R_i} m^{\nu+1}\,dm \nonumber \\
    &= a_0\frac{R_j^{\lambda-\nu}-L_j^{\lambda-\nu}}{\lambda-\nu}\left(R_i^{\nu+2}-L_i^{\nu+2}\right).
    \label{eq:Bij-crossbin-m-integral}
\end{align}
%%%%%%%%%%%%%%
Substituting Eq.~\eqref{eq:Bij-crossbin-m-integral}, together with $\displaystyle\int_{L_i}^{R_i} m\,dm = (R_i^2-L_i^2)/2$, into Eq.~\eqref{eq:Bij-crossbin-preform} yields the explicit form
%%%%%%%%%%%%%%
\begin{equation}
    B_{ij} = \frac{R_i-L_i}{R_j-L_j}\,\frac{2a_0}{R_i^2-L_i^2}\,\dfrac{R_j^{\lambda-\nu}-L_j^{\lambda-\nu}}{\lambda-\nu}\left(R_i^{\nu+2}-L_i^{\nu+2}\right), \qquad i<j.
    \label{eq:Bij-crossbin-final}
\end{equation}
%%%%%%%%%%%%%%

Finally, collecting Eq.~\eqref{eq:Bii-final} and Eq.~\eqref{eq:Bij-crossbin-final} gives the complete explicit form of $B_{ij}$
%%%%%%%%%%%%%%
\begin{align}
    B_{ij} = 
    \begin{cases}
        \dfrac{2a_0}{R_i^2 - L_i^2} \left[ \dfrac{R_i^{\lambda+2} - L_i^{\lambda+2}}{\lambda+2} 
        - L_i^{\nu+2} \, \dfrac{R_i^{\lambda-\nu} - L_i^{\lambda-\nu}}{\lambda-\nu} \right], 
        & i = j, \\[3ex]
        \dfrac{R_i - L_i}{R_j - L_j} \, \dfrac{2a_0}{R_i^2 - L_i^2} \, \dfrac{R_j^{\lambda-\nu} - L_j^{\lambda-\nu}}{\lambda - \nu} \, \left(R_i^{\nu+2} - L_i^{\nu+2}\right),
        & i < j.
    \end{cases}
    \label{eq_frag_Bij_spellout}
\end{align}
%%%%%%%%%%%%%%

%%%%%%%%%%%%%%%%%%%%%%%%%%%%%%%%%%%%%%%%%%%%%%%%%%%%%%%%%%%%%%%%%%%%%
\bibliographystyle{elsarticle-harv} 
\bibliography{Zien_paper_3_bibtex}

\end{document}